\documentstyle[preprint,aps]{revtex}   %  Preprint format
\begin{document}
%\draft

%newcommnads
 
\newcommand{\beqa}{\begin{eqnarray}}
 \newcommand{\beq}{\begin{equation}}
  \newcommand{\eeqa}{\end{eqnarray}}
   \newcommand{\eeq}{\end{equation}}
    \newcommand{\rd}{{\rm d}}
     \newcommand{\e}{{\rm e}}
%\begin{document}

\title{
On the morphological stability of two-dimensional epitaxial islands at
high deposition rates}
\author{
Alberto Pimpinelli} 
\address{LASMEA, Universit\'e Blaise Pascal -- Clermont 2,\\ F-63177 Aubi\`ere
cedex, France.}
\maketitle

\begin{abstract}
The morphological stability of two-dimensional islands 
nucleated on a substrate during vacuum or vapour-phase atom deposition is
investigated. Using simple scaling arguments, it is shown that, contrary to
expectation, dendritic islands may be converted into compact ones by increasing
the deposition rate, provided that the size of the critical nucleus is large
enough. Implications for recent observations of Pt deposition on Pt(111) are
discussed.
\end{abstract}



\hskip1.2truecm vacuum and vapour deposition, Pt(111)

\bigskip\bigskip\bigskip\bigskip
\hskip1.2truecm Correspondence: Tel. (33)~ 4~ 73~40~73~44; FAX (33)~
4~ 73~40~73~40

\hskip4truecm e-mail: Alberto.Pimpinelli@lasmea.univ-bpclermont.fr
\newpage

Since the seminal work by Mullins and Sekerka \cite{muse}, it is known that
compact growth shapes are unstable, in general, when growth takes place by
diffusion in some medium. In particular, scanning tunneling
microscopy (STM) demonstrates the occurrence of fractal two-dimensional (2D)
islands during metal-on-metal epitaxial deposition at low temperature, as
reported by several authors \cite{Hwa,Miche,Bru}. The physical mechanism leading
to ramified---possibly fractal---growth shapes is the same as in the diffusion
limited aggregation (DLA) model of Witten and Sander \cite{wisa}. The deposited
atoms (adatoms) diffuse on the surface, until they meet another adatom or an
island. Any fluctuation of the shape leading to a protrusion
tends to be amplified during growth, since diffusing adatoms find more
easily the protrusion than the flat parts, which lag behind ({\it cf.} the effect
of a point in electrostatics). At low enough temperature the interatomic bond
cannot be broken, and island restructuring by detachment-reattachment of bound
atoms is prevented. If diffusion of bound atoms along the island edge is also
hindered, as it is the case for (111) metal faces, fractal shapes occur. This
instability is intrinsic of diffusion limited growth at low temperature. Indeed,
non-compact or, generally speaking, dendritic growth shapes appear at any 
temperature, provided that smoothing processes, e.g. edge diffusion, are not able
to effectively counterbalance the intrinsic instability \cite{miche2}.  

On the theoretical side, a detailed analysis of the morphological stability of a
2D island has been performed by Avignon and Chakraverty \cite{avcha} for small
adatom supersaturations. The regime of high supersaturation, when the size of a
critical nucleus is 1 \cite{nuc} and classical fractals appear, has been
investigated by Pimpinelli et al. \cite{pim1}, with an emphasis on the
stabilizing action of several matter-transport processes (edge diffusion,
detachment-attachment). The same situation has been subsequently addressed with
Kinetic Monte Carlo simulations by Bartelt and Evans
\cite{baev}, who studied the shape transitions of a single island, and by Bales
and Chrzan \cite{bach}, who studied the compact-fractal transition during epitaxial
deposition. Both works assume edge diffusion as the smoothing mechanism, and
forbid atom detachment from the island. Their results are in good agreement with
the analytic computations in \cite{pim1}. Also, the dependence of the fractal
instability on the geometry of the substrate has been investigated by Zhang et al.
\cite{zha}.

One of the results of such studies is that, not surprisingly, the fractal
instability is favoured by a high deposition rate: if the time  interval between
the arrival of two adatoms onto the island edge is short, edge diffusion cannot
exert its smoothing action, because before the first atom has had the time to
search for a site of high coordination---which is a condition for a compact
shape---it is blocked by the second incoming atom---and limited edge mobility
clearly favours the instability. It appears thus obvious that such
morphological instabilities are due to the far-from-equilibrium nature of the
deposition process, and that going farther out of equilibrium (growing faster)
can only make things worst. 

However, during atom deposition on a substrate, many islands are simultaneously
nucleated, which then grow laterally and compete for capturing additional incoming
atoms. As a result, each island has a ``capture area'' limited by its neighbours
and its advancing edge slows down in approaching the capture area of other islands.
A sort of repulsion between island edges appears, which acts as an indirect
shape-smoothing mechanism. In general, increasing the deposition rate increases
the island density, and thus the effective repulsion. Indeed, I will show that,
under appropriate conditions, increasing the deposition rate makes the islands
more compact.

Let me assess the picture by recalling known results. Note that throughout this
Letter, lenghts are measured in units of the interatomic distance $a$, so that
$a=1$.

The evolution of the island morphology is determined by the balance between
destabilizing and smoothing processes. Since kinetics is involved, it is a matter
of timescales. Calling $F$ the deposition rate, and $\ell^2$ the average size
of the capture area of an island, the average number of atoms reaching the island
per unit time is $\Phi = F\ell^2$. The time elapsed between two
atoms successively hitting the same site on the island edge is then $\tau =
1/\Phi$. For the instability to be checked, the time $\tau$ must be larger than
the average time taken by an atom to diffuse along the island edge. If $D_e$ is
the edge diffusion constant, and $L^2$ the average island size, there is no
instability as long as $\tau> L^2/D_e$. Thus, the instability takes place when 
$L^2$ exceeds the critical value $L_c^2$ given by \cite{pim1,baev}
\beq
\label{1}
L_c^2\approx D_e/\Phi \;.
\eeq

Therefore, the instability has in this case a threshold which decreases with
increasing $F$. Note at this point that confusion should be avoided
between the critical size for the appearance of the instability, $L_c^2$, and the
size of the critical nucleus, $n$. The latter is the size beyond which an island
grows continuously without net loss of atoms. Aggregates smaller than the size
$n$ may also shrink to smaller sizes (see note \cite{nuc}). The DLA-type
situation corresponds to $n=1$ and very low temperature. At higher temperature,
the possibility of atom emission from kinks has to be accounted for, and the edge
diffusion constant $D_e$ must be replaced with a kinetic coefficient $\Gamma_e$,
which also depends on the formation energy of a kink, and thus on the step line
tension (see Refs. \cite{muse,pim1} and point (iii) below.)

When just one island is present, $\ell^2$ is just the substrate area, and
$\Phi$ is just the number of deposited atoms per unit time. Equation (\ref{1})
then states that $L_c^2$ varies as $1/F$. This is an extreme case. Indeed, when
many islands are nucleated on the surface, as it is usually the case in epitaxial
deposition, the capture area $\ell^2$ has to be computed self-consistently by
accounting for the competing presence of the other islands. This task needs
considering the details of island nucleation. It has been performed in a number of
papers \cite{sto,vena,vill}, and the result is the following: $\ell^2$ is
proportional to a power of the ratio $D_s/F$, where $D_s$ is the surface adatom
diffusion constant, the power being a function of the critical nucleus size $n$:
\beq
\label{2}
\ell^2 \approx (D_s/F)^{n/(2+n)}\;.
\eeq

Since each island sits, on average, at the center of a region whose area is
$\ell^2$, the average island density is $N=1/\ell^2$. Hence, the average island
size $L^2$ and the total deposited matter---the surface coverage $\theta=Ft$
---are related {\it via} $L^2N=\theta$, or
\beq
\label{3}
L^2=\theta\ell^2
\eeq

Using $\Phi=F\ell^2$, and equations (\ref{1}) and (\ref{3}) the critical island
size $L_c^2$ can be converted into a critical coverage $\theta_c$ as follows
\beq
\theta_c\approx L_c^2/\ell^2\approx ({\Gamma_e}/{D_s})\;\frac{D_s}{F\ell^4}\;.
\eeq
Replacing $\ell^2$ with its expression (\ref{2}) yields 
\beq
\label{4}
\theta_c\approx({\Gamma_e}/{D_s})\left({F}/{D_s}\right)^{(n-2)/(n+2)}\;.
\eeq
This result is new.

When $n=1$, equation (\ref{4}) becomes \cite{pim1}
\beq
\label{5}
\theta_c\approx({\Gamma_e}/{D_s})\left({D_s}/{F}\right)^{1/3}\;.
\eeq
Equation (\ref{5}) agrees well with Bales and Chrzan's simulations \cite{bach}. It
shows that for $n=1$, the critical coverage for the appearance of the
instability decreases with increasing $F$ as expected. However, we immediately see that,
contrary to expectation, when $n>2$ the opposite is true: the critical coverage in
(\ref{4}) {\it increases} with increasing $F$. The increased growth rate stabilizes the
compact shape!

Other smoothing mechanisms can be investigated by this approach. One only
needs to know for each healing process, the typical relaxation time of a
perturbation of wavevector $q$. These timescales are contained in Ref.
\cite{pim2}. Instead, I rely here on an argument {\it \`a la} Mullins and Sekerka
\cite{muse,avcha,pim1}, which makes use of more macroscopic concepts as the curvature of
the island profile, and the excess chemical potential of the curved parts
(Gibbs-Thomson relation \cite{vill}). Indeed, consider a circular island of
radius $R$. Let $\delta_q$ be the amplitude of a deformation of the circular shape whose
wavevector is $q$. Then, its rate of variation, $\dot\delta_q$, results from the
destabilizing diffusion contribution \cite{muse,avcha,pim1}, 
\beq
\label{dest}
D_s\nabla c\cdot {\bf n}\approx v_{\rm step}q\delta_q\approx qF\ell^2/R~\delta_q\;,
\eeq 
where $c$ is the adatom density, ${\bf n}$ the local step edge
normal and $v_{\rm step}\approx F\ell^2/R$ is the advance velocity of the island
edge step, plus the various smoothing terms. Three of them will be considered
here: i) exchange of atoms between the island and a 2D adatom gas on the
substrate; ii) atom detachment, diffusion on the surface and
reattachment; iii) edge diffusion again, for completeness. Smoothing
occurs because, according to the Gibbs-Thomson relation \cite{vill}, an increase
in the curvature of an interface raises the chemical potential of the curved
part, by a quantity $\delta\mu\approx \tilde\gamma q^2\delta_q$, where
$\tilde\gamma$ is the interface stiffness, and $q$ the wavevector of the
deformation. The density of atoms near the interface is accordingly increased
with respect to the equilibrium density $c_{\rm eq}$:
\beq
\label{6}
c_{\rm excess}=c_{\rm eq}\exp({\beta~\delta\mu})\approx c_{\rm
eq}(1+\beta\tilde\gamma q^2\delta_q )\;,
\eeq
where $1/\beta=k_BT$.

i) When the excess atoms detach from the island and exchange with a 2D adatom
gas on the substrate at rate $k_{\rm exch}$, the smoothing term has the form
\cite{pim2} $k_{\rm exch} c_{\rm excess}\approx \Gamma_{\rm exch} q^2 \delta_q$, where
(\ref{6}) has been used, and  $\Gamma_{\rm exch}=k_{\rm exch} c_{\rm
eq}\beta\tilde\gamma$. It is then straightforward to show in this case that no
critical island size must be exceeded for the instability to appear.
Any size is unstable, provided that $F/D_s > (D_s/\Gamma_{ex})^{(n+2)/2}$. If the
opposite is true, no size is unstable. When this smoothing process dominates, a
``typical'' situation occurs, in which increasing $F$ is destabilizing. What
smoothing process dominates is a system-dependent matter, of course.

ii) When the excess atoms detach from the island and surface diffuse towards
places of lesser curvature, the smoothing term has the form \cite{pim1}
\beq
\label{8}
D_s \nabla c_{\rm excess}\cdot {\bf n}\approx \Gamma_s q \times q^2\delta_q
\eeq
where (\ref{6}) has been used, and  $\Gamma_s=D_s c_{\rm eq}\beta\tilde\gamma$.
Equating (\ref{8}) to (\ref{dest}) the instability threshold $q_c$ is found,
\beq
Rq_c \approx (F\ell^2R/\Gamma_s)^{1/2} \;.
\eeq
The wavevelength $1/q$ of a shape perturbation cannot be arbitrarily
large. Indeed, if $R$ is the island radius one has $1/q\leq R$. The critical
radius $R_c$ can then be found by letting $q=1/R$, which yields $R_c\approx
\Gamma_s/(F\ell^2)$, and eventually leads to 
\beq
\theta_c\approx({\Gamma_s}/{D_s})^2\;\left({F}/{D_s}\right)^{(n-4)/(n+2)}\;.
\eeq
when (\ref{2}) and (\ref{3}) are used. The qualitative result is the same as
that of (\ref{4}), i.e. when $n$ is large enough ($n>4$ here), increasing $F$
increases the stability of the compact shape.

iii) When excess atoms diffuse along the interface, they give rise to a smoothing
term of the form \cite{pim1}
\beq
\label{7}
D_e \nabla^2 c_{\rm excess}\approx \Gamma_e q^2 \times q^2\delta_q
\eeq 
where (\ref{6}) has been used, and $\Gamma_e=D_ec_{\rm eq}\beta\tilde\gamma$.
Equating (\ref{7}) to (\ref{dest}) the instability threshold $q_c$ is found,
\beq
Rq_c \approx (F\ell^2R^2/\Gamma_e)^{1/3} \;.
\eeq
The critical radius $R_c$ can then be found by letting $q=1/R$, which
yields (\ref{1}), and eventually leads to (\ref{4}) when (\ref{2}) and (\ref{3})
are used. The reader can check that the alternative method based on Ref.
\cite{pim2} gives the same results in all cases. 

The interesting point is that the scenario sketched here appears
to have been observed. Michely, Hohage and Comsa \cite{miche2} report that the
triangular islands formed during Pt deposition on Pt(111) at 445 K and
$F=2.1\times 10^{-4}$ ML/s, are morphologically unstable. Under these
conditions, they exhibit dendritic shapes at a coverage $\theta=0.1$ monolayer
(ML). On {\it increasing} the flux by two orders of magnitude and leaving all
other conditions unchanged, the islands appear to be quite compact. Stability is
indeed obtained by depositing faster! A quantitative comparison of the
qualitative arguments given in the present Letter and experiment is obviously out
of the question. Pt(111) is known as a system with complicated behaviour
\cite{Miche,miche2}, which only detailed kinetic Monte Carlo simulations can hope
to grasp \cite{liu,hoha}. I could only venture to suggest that for such a system
at a relatively high substrate temperature, the critical nucleus size is likely
to be large, possibly $n=6$ \cite{note}, so that both mechanisms (ii)
and (iii) would predict the observed stabilization at increasing deposition
rate, if they happen to dictate step smoothing on Pt(111).


\begin{thebibliography}{99}
\bibitem{nuc}A size of the critical nucleus, $n$, equal to 1 means that a bound
pair of atoms (dimer) will not break apart---or decay---into two free adatoms.
Thus, $n=2$ means that a trimer will not decay, etc.
\end{thebibliography}
\end{document}